\documentclass[10pt]{article}

\usepackage[utf8]{inputenc}
\usepackage{amsmath}
\usepackage{amsfonts}
\usepackage{bbm}
\usepackage{amssymb}
\usepackage{makeidx}
\usepackage{graphicx}
\usepackage{verbatim}
\usepackage{wrapfig}
\usepackage{xcolor}
\usepackage{multicol}
\usepackage{comment}
\usepackage{setspace}
\usepackage{soul}
\usepackage{makeidx}
\usepackage{braket}
\usepackage{multirow}
\usepackage[font=small,labelfont=it]{caption}
\usepackage[font=scriptsize]{caption}
\usepackage{blindtext}
\usepackage[hidelinks]{hyperref}
\hypersetup{
        colorlinks=true,
        linkcolor=blue,
        citecolor=magenta,
}

\newcommand{\al}{\alpha}
\newcommand{\be}{\beta}

\renewcommand{\be}{\beta}
\newcommand{\ga}{\gamma}
\newcommand{\de}{\delta}

\newcommand{\bA}{\bar{A}}
\newcommand{\bps}{\bar{\psi}}
\newcommand{\bp}{\bar{\partial}}
\newcommand{\pp}{\partial^+}

\renewcommand{\>}{{\rangle}}
\newcommand{\<}{{\langle}}

\renewcommand{\[}{{\langle\!\langle}}

\newcommand{\sfrac}[2]{{\textstyle\frac{#1}{#2}}}

\newcommand{\pa}{\partial}
\newcommand{\ti}{{\times}}
\newcommand{\tr}{{\mathrm{tr}}}
\newcommand{\im}{{\mathrm{i}}}

\newcommand{\diff}{{\mathrm{d}}}

\newcommand{\unity}{\mathbbm{1}}

\newcommand{\beq}{\begin{equation}}
\newcommand{\eeq}{\end{equation}}
\newcommand{\eq}{\end{equation}}
\newcommand{\bea}{\begin{eqnarray}}
\newcommand{\eea}{\end{eqnarray}}
\newcommand{\with}{{\quad{\rm with}\quad}}

\renewcommand{\and}{{\quad{\rm and}\quad}}
\newcommand{\und}{{\qquad{\rm and}\qquad}}
\renewcommand{\=}{\ =\ }

\newcommand{\nn}{\nonumber}

\advance \textheight by \topskip
\makeatletter
\@addtoreset{equation}{section}

\makeatother

\begin{document}

\begin{titlepage}
\setcounter{page}{0}
\begin{flushright}
{\small $\,$}
\end{flushright}
\vskip 1cm
\centerline{\huge{\bf{Nicolai maps for super Yang--Mills}}}
\vskip 0.4cm
\centerline{\huge{\bf{on the light cone}}}
\vskip 1.5cm
\begin{center}
\begingroup\scshape\Large
Olaf Lechtenfeld
\endgroup
\end{center}
\vskip 0.5cm
\centerline{
\it {Institut f\"ur Theoretische Physik and Riemann Center for Geometry and Physics}}
\centerline {\it {Leibniz Universit\"at Hannover, Appelstrasse 2, 30167 Hannover, Germany}}
\vskip 1.5cm
\centerline{\bf {Abstract}}
\vskip .5cm
\noindent
We construct Nicolai maps for supersymmetric Yang--Mills theory in four and ten spacetime dimensions 
in the light-cone gauge, where the elimination of non-propagating degrees of freedom causes nonlocal 
and four-fermi interactions in the Lagrangian. The presence of the latter used to be an obstruction
to the Nicolai map, which has recently been overcome at the price of quantum corrections to the map. 
No gauge-fixing or ghost terms arise in this formulation, since only physical transverse degrees of freedom occur.
We present an explicit form of the Nicolai map to second order in the gauge coupling.
In four dimensions, a `chiral' choice of the map leaves one of the two transverse gauge-field modes invariant,
which forces the classical part of the map (on the other mode) to become a polynomial 
(quadratic in the gauge coupling, cubic in the gauge field)! 
In the power series expansion for the ten-dimensional map however, cancellations at each order in the coupling 
are systematic but incomplete, still leaving an infinite power series for the Nicolai map
(on all eight transverse modes). 
Nevertheless, the existence of a polynomial variant is conceivable, 
also for the maximal ${\cal N}{=}\,4$ theory in four dimensions.
\vfill
\end{titlepage}

\section{Introduction and summary}

\noindent
The Nicolai map~$T$~\cite{Nic1,Nic2,Nic3} (for a more recent review, see~\cite{Lprague})
relates a supersymmetric Yang--Mills theory in spacetime dimension~$D$ at different values of its coupling constant.
In particular, it reduces the expectation value of any functional~$Y$ built from the gauge fields~$A$ 
at coupling~$g$ to a free-field ($g{=}0$) correlator of the inversely transformed functional~$T^{-1}Y$,\footnote{
The proper normalization $\langle1\rangle_g=1$ follows from the vanishing of the vacuum energy.}
\begin{equation} \label{Tdef}
\bigl\langle Y[A] \bigr\rangle_g 
\= \bigl\langle (T_g^{-1}Y)[A] \bigr\rangle_0
\= \bigl\langle Y[T_g^{-1}A] \bigr\rangle_0\ .
\end{equation}
Here, the subscript on the correlator and on the symbol of the map~$T_g:A\mapsto A'[g,A]=T_gA$ 
indicates the value of the coupling,
and the distributivity $T(A_1A_2)=(TA_1) \, (TA_2)$ is used in the second equality.
In the expectation values of~\eqref{Tdef} all anticommuting degrees of freedom (gaugini~$\psi$ and ghosts~$c$)
have been integrated out.\footnote{
Auxiliary fields are also eliminated, after all supersymmetry variations have been performed (see below).}
Therefore, the Nicolai map acts in a nonlocal bosonic theory with an action
\begin{equation}
S_g[A] \= S^{(0)}_g[A] + {\textstyle\sum}_{r=1}^\infty \hbar^r S_g^{(r)}[A]\ .
\end{equation}
Here, the local $S^{(0)}_g$ is the bosonic part of the original supersymmetric action,
\begin{equation}
S^{(0)}_g[A] \= S_{\textrm{\tiny SUSY}}[A,\psi{=}0,c{=}0] \qquad\textrm{with}\qquad
S_{\textrm{\tiny SUSY}}[A,\psi,c] \= \smallint\!\diff^D\!x\ 
\bigl\{ {\cal L}^{\textrm{inv}}[A,\psi]+{\cal L}^{\textrm{gf}}[A,c]\bigr\}\ ,
\end{equation}
composed of a gauge-invariant (inv) and a gauge-fixing (gf) part.
The nonlocal $\exp\{\frac{\im}{\hbar}\sum_r\hbar^r S_g^{(r)}\}$ is produced by the path integral 
over the anticommuting fields in the partition function ($r$ is the number of their loops).
Thus, the Nicolai map connects a super Yang--Mills theory with  a gauge group of dimension~$d$
to its weak-coupling limit, which is a $\text{U}(1)^d$ abelian theory.

If the anticommuting fields $(\psi,c)$ appear in powers higher than quadratic in the supersymmetric action, 
then $S_g^{(r)}\neq0$ for $r{>}1$, and the Nicolai map is a formal power series not only in~$g$ 
but also in~$\hbar$~\cite{CLR},
\begin{equation} \label{quantmap}
T_gA \= T_g^{(0)}\!A \ + {\textstyle\sum}_{r=1}^\infty \hbar^r T_g^{(r)}\!A\ .
\end{equation}
Changing path-integral variables $A\mapsto T_gA$ on the right-hand side of~\eqref{Tdef}
leads to the identity
\begin{equation} \label{matchingnew}
S^{(0)}_0[T_gA]\ +\ \sum_r \hbar^r S_0^{(r)}
\ -\ \im\hbar\,\tr\ln\sfrac{\delta T_gA}{\delta A}
\= S^{(0)}_g[A]\ +\ \sum_r \hbar^r S_g^{(r)}[A]\ ,
\end{equation}
where $S_0^{(r)}$ are constants for~$r{>}0$.
Inserting~\eqref{quantmap} into~\eqref{matchingnew} and collecting equal powers in~$\hbar$
one finds an infinite hierarchy of `Nicolai-map conditions'~\cite{CLR}. 
The leading two correspond to the tree-level and one-loop contributions and take the form
\begin{equation} \label{matching01}
S^{(0)}_0[T_g^{(0)}\!A] \= S^{(0)}_g[A] \quad\und\quad
S^{(0)}_0[T_gA]\big|_{O(\hbar)} + S^{(1)}_0
- \im\,\tr\ln\sfrac{\delta T_g^{(0)}\!A}{\delta A} \= S_g^{(1)}[A] \ .
\end{equation}
The first one is called the `free-action condition', while the second one is named
the `determinant-matching condition'.
There are further conditions, each one balancing expressions of a fixed loop order.

For any off-shell supersymmetric gauge theory, a universal formula provides a formal power series expansion
(in~$g$) for $T_gA$ and its inverse in terms of the `coupling flow operator'~\cite{DL1,L1,L2}
\begin{equation}
R_g[A] \= \smallint\!\diff z\ \bigl(\pa_g T_g^{-1} \circ T_g \bigr)\,A(z)\,\frac{\delta}{\delta A(z)}\ ,
\end{equation}
where `$z$' stands for all coordinates~$x$ and indices the gauge fields carry.
This functional differential operator rules the infinitesimal Nicolai map~\cite{L1},
\begin{equation} \label{localflow}
\pa_g \bigl\langle Y[A] \bigr\rangle_g \= \bigl\langle \bigl( \pa_g + R_g[A] \bigr) Y[A] \bigr\rangle_g\ .
\end{equation}
For its construction it is crucial that, in off-shell supersymmetric theories,
there exists at least one anticommuting functional~${\cal M}_\alpha$ such that
\begin{equation} \label{defDelta}
{\cal L}^{\textrm{inv}}[A,\psi] \= \delta_\alpha {\cal M}_\alpha[A,\psi] \ +\ \textrm{total derivative}\ ,
\end{equation}
where $\alpha$ is a spinor index and $\delta_\alpha$ denotes a supervariation.
The coupling flow operator can then be constructed by combining $\pa_g{\cal M}_\alpha$ with a supervariation
$\delta_\alpha=\smallint(\de_\alpha A)\frac{\de}{\de A}$ on the gauge fields, and averaging over gaugini and ghosts.
The latter generates certain fermionic correlators~$\langle\ldots\rangle_{\psi,c}$ in the expression for~$R_g$,
whose $g$~dependence is essential for the map.
The (finite-flow) Nicolai map is finally obtained as a $g$-ordered exponential of ${-}\!\!\int_0^g\diff{g'}\,R_{g'}$,
requiring the $R_g$ action to be iterated on~$A$~\cite{LR1}.

Actually, we have been oversimplifying somewhat. The non-commutation of $\pa_g$ and $\de_\alpha$ for nonlinear
supersymmetry transformations as well as the gauge-fixing part of~$S_{\textrm{\tiny SUSY}}$ substantially complicates 
the analysis for supersymmetric Yang--Mills theory, at least in covariant gauges~\cite{L1}.
Nevertheless, Nicolai maps have been constructed, mostly in the Landau gauge, for super Yang--Mills theory in
$D=3,4,6$ and~$10$~\cite{ALMNPP,MN,LR2,LR4}.

In this work we avoid all these difficulties by going to the light-cone gauge and eliminating the unphysical
(non-propagating) degrees of freedom~\cite{HRT,BGS}. This introduces some nonlocalities into~$S_{\textrm{\tiny SUSY}}$
but renders half of the off-shell supersymmetries linear, and no gauge-fixing or ghost part is required in the action.
As a consequence, the Nicolai-map construction proceeds in the same (easy) fashion as for chiral theories~\cite{Nic3,Lprague}.
The only novel complication is a quartic gaugino interaction, which falls outside the scope of the traditional Nicolai map.
Recently, however, the existence and construction of the map has been generalized to arbitrary fermionic interactions,
at the expense of allowing for quantum corrections to the (formerly purely classical) Nicolai map~\cite{CLR}.
There have been previous attempts and proposals for a super Yang--Mills Nicolai map in the light-cone gauge,
early ones~\cite{dAFFV} and more recently~\cite{MN,LR2}, but these kept unphysical degrees of freedom in the
gauge-fixed local super Yang--Mills action, while here we eliminated these at the price of some nonlocality.

For ${\cal N}{=}\,1$ supersymmetric pure Yang--Mills theory in both $D{=}4$ and $D{=}10$ we explicitly construct 
the Nicolai map~$T_gA$ to second order in the gauge coupling~$g$, see \eqref{T4} and~\eqref{T10}. 
We also verify the free-action condition in either case.
It is seen that $r$-loop quantum contribution~$T_g^{(r)}\!A$ first occurs at order~$g^{2r}$. 
Between the iterated actions of the coupling flow operator, substantial cancellations occur at each order of~$g$
in the construction, but only in $D{=}4$ do these completely remove the classical part~$T_g^{(0)}\!A$ beyond~$O(g^2)$,
rendering the classical Nicolai map quadratic in~$g$ and cubic in~$A$ for this case!
This parallels the collapse of the map to a polynomial one for supersymmetric quantum mechanics and for ${\cal N}{=}\,2$
Wess--Zumino models in~$D{=}2$~\cite{Nic3,L1,L2,LR3}. In all these cases an additional freedom to select a `chiral' version
of the map (based on just a quarter of the original supersymmetries) is essential. In $D{=}4$ Yang--Mills theory,
this `chiral' map affects only one (complex) combination of the two transversal gauge-field components, leaving the 
complementary combination inert! The free-action condition then immediately rules out any contribution to $T_g^{(0)}\!A$
in $g$-power higher than the bosonic action. As usual, formally inverting the power series to obtain $T^{-1}_gA$ even classically
cannot terminate and will reproduce via~\eqref{Tdef} the perturbative expansion of any correlator, though in a non-standard form.

There are several ways in which the work presented here can be further expanded or generalized.
First, the quantum contributions $T_g^{(r)}\!A$ may be worked out for $r{=}1$ or higher, and the corresponding
`Nicolai-map loop conditions' checked, with an appropriate regularization procedure. It will be curious to see
whether the $D{=}4$ map also collapses at the quantum level. 
Second, Ward identites of the form $\langle Y[T_gA]\rangle_g = \langle Y[A]\rangle_0$ following directly from~\eqref{Tdef}
can be exploited in particular for $D{=}4$ super Yang--Mills theory.
Third, dimensional reduction of~\eqref{T10} to $D{=}4$ produces the Nicolai map for the complex transverse gauge potential~$(A,\bA)$
and the six adjoint scalars~$C^{k\ell}$ of the ${\cal N}{=}\,4$ super Yang--Mills theory in the light-cone frame~\cite{BLN,Man}.
Simplifications are not apparent in this reduction, but cannot be excluded for a clever recombination of the field variables
or some kind of `chiral' variant of the map.
Finally, one would of course like to see real computational applications of the Nicolai map. 
First steps towards this goal have been made in~\cite{DL2,NP,Mal}.
%

\section{The map in four dimensions}

\noindent
Four-dimensional ${\cal N}{=}\,1$ super Yang--Mills theory in the light-cone gauge~$A^+{=}\,0$ 
can be presented, after elimination of all non-propagating auxiliary degrees of freedom, by the Lagrangian~\cite{HRT}
\begin{equation} \label{lag4}
\begin{aligned}
{\cal L} \ &=\ -\bA\cdot\Box\,A + \im\,\psi\cdot\tfrac{\Box}{\pp}\bps \\[4pt]
&\ -\,g\,\bigl\{ [(\pp\!\bA)\ti A+\im\,\bps\ti\psi]\cdot\tfrac{\bp\ }{\pp}A
+[(\pp\!A)\ti\bA+\im\,\psi\ti\bps]\cdot\tfrac{\pa\ }{\pp}\bA
-\im\,[A\ti\psi]\cdot\tfrac{\bp\ }{\pp}\bps -\im\,[\bA\ti\bps]\cdot\tfrac{\pa\ }{\pp}\psi \bigr\} \\[4pt]
&\ +\,\tfrac12 g^2 \bigl\{ [(\pp\!\bA)\ti A+\im\,\bps\ti\psi]\cdot\tfrac1\pp\tfrac1\pp[(\pp\!A)\ti\bA+\im\,\psi\ti\bps]
+\im\,[\bA\ti\bps]\cdot\tfrac1\pp[A\ti\psi] \bigr\} 
\end{aligned}
\end{equation}
with a complex transversal derivative $\pa{=}\tfrac1{\sqrt{2}}(\pa_1{+}\im\pa_2)$,
for a complex scalar field $A{=}\tfrac1{\sqrt{2}}(A_1{+}\im A_2)$ 
and a complex Grassmann-valued field $\psi{=}\tfrac1{\sqrt{2}}(\psi_1{+}\im\psi_2)$,
both in the adjoint representation of the gauge group, say SU($N$), with gauge coupling~$g$.
The color structure (adjoint index $a$) is hidden in the notation
\begin{equation}
B\cdot C := B^a C^a \und (B{\times}C)^a := f^{abc} B^b C^c
\end{equation}
with totally antisymmetric structure constants $f^{abc}$, and square brackets indicate the order of color multiplication.
Our convention is that any derivative or its inverse acts on everything to its right, unless limited by round rackets it resides in between.
Finally, the nonlocal $\tfrac1{\pp}$ is understood as an antisymmetric integral kernel, {\it i.e.}
\begin{equation}
\smallint\diff x\ B\,(\tfrac1{\pp}C) = - \smallint\diff x\ (\tfrac1{\pp}B)\,C
\und \tfrac1{\pp}\,\textrm{constant}= 0\ .
\end{equation}

The action $S=\smallint\diff^{4}x\,{\cal L}$ is  manifestly invariant under half of the original 4 supersymmetries,
\begin{equation} \label{susy4}
\begin{aligned}
&\bar\de A = \im\,\bps\ ,\quad 
\bar\de\bA = 0\ ,\quad
\bar\de\psi = -\pp\!\bA\ ,\quad
\bar\de\bps = 0\ , \\
&\de A = 0\ ,\quad 
\de\bA = \im\,\psi\ ,\quad 
\de\psi = 0\ ,\quad 
\de\bps = -\pp\!A\ ,
\end{aligned}
\end{equation}
which are linearly realized.
For this reason, there exists a light-cone superspace in which $(A,\psi)$ are combined into a chiral superfield~\cite{HRT}.
Therefore, we are assured of the existence of a complex spinor ${\cal M}$ which generates the Lagrangian~\eqref{lag4} via
a supervariation,
\begin{equation}
{\cal L} \= \bar\de\,{\cal M} \ + \textrm{total derivative} \= \de\,\bar{\cal M} \ + \textrm{total derivative}\ .
\end{equation}
In general, any linear combination of $\bar\de{\cal M}$ and $\de\bar{\cal M}$ (with total weight one) 
can be used to express~$\cal L$ but, since their difference is a total derivative, we restrict ourselves to
a simple (``chiral'') choice employing only~the holomorphic part~$\cal M$.
Furthermore, due to the linearity of~\eqref{susy4}, the supervariation commutes with $\tfrac{\pa}{\pa g}$, and hence
\begin{equation}
\tfrac{\pa}{\pa g} \smallint\diff^{4}x\ {\cal L}(x) \= \bar\de\smallint\diff^{4}x\ \tfrac{\pa}{\pa g}{\cal M}(x)\ .
\end{equation}

Employing $\bar\de\bar\de=0$ and the useful relations
\begin{equation}
\bar\de\,(A\ti\psi) = (\pp\!\bA)\ti A+\im\,\bps\ti\psi \und
\bar\de\,((\pp\!A)\ti\bA+\im\,\psi\ti\bps) = -\im\,\pp\!(\bA\ti\bps)\ ,
\end{equation}
it is easy to find
\begin{equation}
{\cal M} \= A\cdot\tfrac{\Box}{\pp}\psi 
\,-\,g\,\bigl\{ A\cdot\tfrac{\bp\ }{\pp}[A\ti\psi]+\bA\cdot\tfrac{\pa\ }{\pp}[A\ti\psi]+[A\ti\bA]\cdot\tfrac{\pa\ }{\pp}\psi\bigr\}
\,+\,\tfrac12 g^2 \bigl\{ [(\pp\!A)\ti\bA+\im\,\psi\ti\bps]\cdot\tfrac1\pp\tfrac1\pp[A\ti\psi] \bigr\}\ .
\end{equation}
This expression enters the gauge-coupling flow operator,
\begin{equation}
\begin{aligned} \label{Rg4}
R_g &\= \im\smallint\diff^{4}x\ \< \tfrac{\pa}{\pa g}{\cal M}(x)\ \bar\de \>_\psi
\= -\smallint\diff^{4}x \smallint\diff^{4}y\ \< \tfrac{\pa}{\pa g}{\cal M}(x)\ \bps(y) \>_\psi\ \tfrac{\de}{\de A(y)} \\[4pt]
&\= \smallint\diff^{4}x \smallint\diff^{4}y\ \bigl\{
A(x)\ti\tfrac{\bp\ }{\pp}A(x)+\bA(x)\ti\tfrac{\pa\ }{\pp}A(x)+A(x)\ti\bA(x)\,\tfrac{\pa\ }{\pp}\bigr\}
\cdot\<\psi(x)\ \bps(y)\>_\psi\cdot\tfrac{\de}{\de A(y)}\\[2pt]
&\ -g\smallint\diff^{4}x \smallint\diff^{4}y\ 
(\pp\!A(x))\ti\bA(x)\cdot\tfrac1\pp\tfrac1\pp A(x)\ti\,\<\psi(x)\ \bps(y)\>_\psi\cdot\tfrac{\de}{\de A(y)}\\[2pt]
&\ -g\smallint\diff^{4}x \smallint\diff^{4}y\ 
\< \im\,\psi(x)\ti\bps(x)\cdot\tfrac1\pp\tfrac1\pp A(x)\ti\,\psi(x)\ \bps(y)\>_\psi\cdot\tfrac{\de}{\de A(y)}\ ,
\end{aligned}
\end{equation}
where we dropped the square brackets since the color structure is unambiguous, 
the (inverse) derivatives act only on functions of~$x$ to the right like
\begin{equation}
\tfrac{\pa\ }{\pp}\,B(x)\,C(y) \ :=\ (\tfrac{\pa\ }{\pp} B)(x)\,C(y)  \und
\tfrac{1}{\pp}\,B(x)\,C(y) \ :=\ (\tfrac{1}{\pp} B)(x)\,C(y)\ ,
\end{equation}
and $\<\ldots\>_\psi$ denotes the result of integrating out the gaugini, {\it i.e.} the fermionic path integral.
While the first two lines of~\eqref{Rg4} contain only fermionic bilinears $\<\psi(x)\,\psi(y)\>_\psi$
and thus produce full gaugino propagators (in the gauge-field background) upon averaging over~$\psi$, \
the last line is quartic in the fermions and hence gives rise to a (fused) four-gaugino amplitude $\<\psi(z)\psi(z)\psi(x)\,\psi(y)\>_\psi$,
which however is suppressed by a (hidden) factor of~$\hbar$.

In order to set up the perturbative expression for the Nicolai map, we develop a power series in~$g$,
\begin{equation} \label{Rexp}
\begin{aligned}
R_g &\= R_g^{(1)} + R_g^{(2)} + R_g^{(3)} \\
&\= r_1^{(1)}\!\ +\ g\,(r_2^{(1)}\!+r_2^{(2)}\!+r_2^{(3)})\ +\ g^2\,(r_3^{(1)}\!+r_3^{(2)}\!+r_3^{(3)})\ +\ 
g^3\,(r_4^{(1)}\!+r_4^{(2)}\!+r_4^{(3)})\ +\ \ldots \\
&\= r_1\ +\ g\,r_2\ +\ g^2\,r_3\ +\ g^3\,r_4\ +\ \ldots \ ,
\end{aligned}
\end{equation}
where the superscript label refers to the first, second, or third line of the expression~\eqref{Rg4}.
We note that $r_1^{(2)}=r_1^{(3)}=0$.
Besides the explicit $g$~dependence in~\eqref{Rg4}, we must take into account the implicit
$g$~dependence of the gaugino two- and four-point functions $\<\ldots\>_\psi$,
\begin{align} \label{psicorr4}
\<\psi^a(x)\ \bps^b(y)\>_\psi &\= \de^{ab}\,\tfrac{\pp}{\Box}(x{-}y) \nn \\
&\ \ + g\,f^{acb}\smallint\diff^{4}z\,\tfrac{\pp}{\Box}(x{-}z)\bigl\{
\bA(z)\tfrac{\pa\ }{\pp} + \tfrac{\bp\ }{\pp}A(z) - (\tfrac{\pa\ }{\pp}\bA(z)) - (\tfrac{\bp\ }{\pp}A(z))
\bigr\}\tfrac{\pp}{\Box}(z{-}y) \nn \\[2pt]
&\ \ + O(g^2)\ ,\\[4pt]
\<(\psi\ti\bps)^c(z)\ \psi^a(x)\ \bps^b(y)\>_\psi &\= 
-f^{cab}\,\tfrac{\pp}{\Box}(z{-}x) \tfrac{\pp}{\Box}(z{-}y) \ + \ O(g)\ . \nn
\end{align}
Inserting \eqref{psicorr4} into \eqref{Rg4} and using $f^{abc}f^{adc}=N\,\de^{bd}$ for SU($N$) we find
\begin{equation} \label{r1}
\begin{aligned} 
r_1^{(1)} &\= \smallint\diff^{4}x \smallint\diff^{4}y\ \bigl\{
A(x)\ti\tfrac{\bp\ }{\pp}A(x)+\bA(x)\ti\tfrac{\pa\ }{\pp}A(x)+A(x)\ti\bA(x)\,\tfrac{\pa\ }{\pp}\bigr\}
\cdot\tfrac{\pp}{\Box}(x{-}y)\,\tfrac{\de}{\de A(y)} \\[2pt]
&\= -\smallint\diff^{4}x \smallint\diff^{4}y\ \bigl\{
A(x)\ti(\tfrac{\bp\ }{\pp}A(x))+A(x)\ti(\tfrac{\pa\ }{\pp}\bA(x))-A(x)\ti\bA(x)\,\tfrac{\pa\ }{\pp}\bigr\}
\cdot\tfrac{\pp}{\Box}(x{-}y)\,\tfrac{\de}{\de A(y)}
\end{aligned}
\end{equation}
and (reinstating $\hbar$)
\begin{equation} \label{r2}
\begin{aligned} 
r_2^{(1)} &\= \smallint\diff^{4}x \smallint\diff^{4}z \smallint\diff^{4}y\ \bigl\{
A(x)\ti\tfrac{\bp\ }{\pp}A(x)+\bA(x)\ti\tfrac{\pa\ }{\pp}A(x)+A(x)\ti\bA(x)\,\tfrac{\pa\ }{\pp}\bigr\}\tfrac{\pp}{\Box}(x{-}z) \\
&\qquad\qquad\qquad\qquad\quad\!
\cdot\bigl\{\bA(z)\tfrac{\pa\ }{\pp}+\tfrac{\bp\ }{\pp}A(z)-(\tfrac{\pa\ }{\pp}\bA(z))-(\tfrac{\bp\ }{\pp}A(z))\bigr\}
\tfrac{\pp}{\Box}(z{-}y)\ti\,\tfrac{\de}{\de A(y)} \ , \\[2pt]
r_2^{(2)} &\= \smallint\diff^{4}x \smallint\diff^{4}y\
\bA(x)\ti(\pp\! A(x))\cdot\tfrac{1}{\pp}\tfrac{1}{\pp}A(x)\tfrac{\pp}{\Box}(x{-}y)
\ti\,\tfrac{\de}{\de A(y)} \ ,\\[2pt]
r_2^{(3)} &\= -N\,\im\hbar \smallint\diff^{4}x \smallint\diff^{4}y\
\tfrac{1}{\pp}\tfrac{1}{\pp} A(x)\cdot\tfrac{\pp}{\Box}(z{-}x)\tfrac{\pp}{\Box}(z{-}y)\big|_{z=x} \,\tfrac{\de}{\de A(y)}\ ,
\end{aligned}
\end{equation}
where $\tfrac{1}{\pp}$ or $\tfrac{\pa\ }{\pp}$ without argument acts to the right on all functions of the first argument appearing,
unless limited by round brackets and before eventually putting $z{=}x$.

With these ingredients, we are able to write down the Nicolai map to order~$g^2$,
\begin{align} \label{Tg4}
(T_gA)(y) &\= A(y)\ -\ \smallint_0^g\!\diff g'\; R_{g'}A(y)\ +\ 
\smallint_0^g\!\diff g'\smallint_0^{g'}\!\diff g''\; R_{g'}R_{g''}A(y)\ -\
\smallint_0^g\!\diff g'\smallint_0^{g'}\!\diff g''\!\smallint_0^{g''}\!\diff g'''\; R_{g'}R_{g''}R_{g'''}A(y)\ +\ \ldots\nn\\
&\=  A(y)\ -\ g\,r_1 A(y)\ -\ \tfrac12g^2 (r_2-r_1r_1)A(y)\ -\ \tfrac16g^3 (2 r_3 -2 r_2r_1 - r_1r_2 + r_1r_1r_1)A(y)\ +\ O(g^4) \nn\\[2pt]
&\= A(y)\ -\ g\,r_1^{(1)}\!A(y)\ -\ \tfrac12g^2 (r_2^{(1)}+r_2^{(2)}+r_2^{(3)}-r_1^{(1)}r_1^{(1)})A(y)\ +\ O(g^3) \\[2pt]
&\ =:\ A(y)\ +\ g\,t_1 A(y)\ +\ g^2\,t_2 A(y)\ +\ g^3\,t_3 A(y)\ +\ O(g^4) \ .\nn
\end{align}
We stress that, because the single $\bar\de$ supersymmetry employed here annihilates $\bA$ in~\eqref{susy4},
the complex conjugate component~$\bA$ is not affected by the Nicolai map: $T_g\bA=\bA$ for our `chiral' choice of the map!
Amazingly, acting with $r_1^{(1)}$ of~\eqref{r1} onto~$r_1^{(1)}A$ produces exactly the structure of $r_2^{(1)}A$ in~\eqref{r2},
and hence these two contributions completely cancel each other, leaving us with
\begin{equation} \label{Tg4a}
(T_gA)(y) \= A(y)\ -\ g\,(r_1^{(1)}\!A)(y)\ -\ \tfrac12g^2 (r_2^{(2+3)}\!A)(y)\ +\ O(g^3)\ .
\end{equation}
We then arrive at the expression
\begin{equation} \label{T4}
\begin{aligned}
(T_gA)(y) &\= (T_g^{(0)}\!A)(y)\ +\ \hbar\,(T_g^{(1)}\!A)(y)\ +\ \hbar^2(T_g^{(0)}\!A)(y)\ +\ O(\hbar^3) \\[4pt]
&\= A(y)\ -\ g\, \smallint\diff^{4}x\ \bigl\{
A(x)\ti\tfrac{\bp\ }{\pp}A(x)+\bA(x)\ti\tfrac{\pa\ }{\pp}A(x)+A(x)\ti\bA(x)\,\tfrac{\pa\ }{\pp}\bigr\} \tfrac{\pp}{\Box}(x{-}y) \\[2pt]
&\ \ -\ \tfrac12 g^2 \smallint\diff^{4}x\ \bigl[\bA(x)\ti(\pp\!A(x))\bigr]\ti\tfrac{1}{\pp}\tfrac{1}{\pp}A(x) \tfrac{\pp}{\Box}(x{-}y) \\[2pt]
&\ \ +\ \tfrac{\im}{2}\hbar N g^2 \smallint\diff^{4}x\ \tfrac{1}{\pp}\tfrac{1}{\pp} A(x)
\tfrac{\pp}{\Box}(z{-}x)\tfrac{\pp}{\Box}(z{-}y)\big|_{z=x} \ +\ O(g^3) \ .
\end{aligned}
\end{equation}
The first two lines belong to the classical map (in fact, saturate it, as we shall prove), 
while the third line is the leading piece of the first quantum correction~$\hbar\,T_g^{(1)}\!A$.

The observed cancellation at $O(g^2)$ generalizes to higher orders,  
at least for the classical part $T_g^{(0)}\!A=T_gA\big|_{\hbar=0}$.
One finds that
\begin{equation}
r_{s\phantom{1}}^{(1+2)}r_1^{(1)}A\ +\ r_{s-1}^{(1+2)}r_2^{(2)}A \= r_{s+1}^{(1+2)}A \qquad\forall s\ge1 \ ,
\end{equation}
with $r_0^{(1+2)}=1$ and $r_1^{(2)}=0$.
Consulting~\cite{LR1} for the $g$ expansion of the Nicolai map in terms of the $r_i$ at all orders we see that, 
as a consequence, all contributions from the perturbative expansion of the full gaugino propagator are cancelled,
leaving in the classical map only the terms from a single action of $R_g$ with the free propagator inserted,
\begin{equation} \label{T4cl}
\begin{aligned}
(T_g^{(0)}\!A)(y) 
&\= A(y)\ -\ g\,(R_0 A)(y)\ -\ \tfrac12g^2\,(R'_0 A)(y)\big|_{\hbar=0} \\[2pt]
&\= A(y)\ -\ g\,(r_1^{(1)}\!A)(y)\ -\ \tfrac12g^2 (r_2^{(2)}\!A)(y)\ ,
\end{aligned}
\end{equation}
with $R'_g\equiv\pa_gR_g$.
In other words, the $O(g^3)$ in \eqref{Tg4a} and~\eqref{T4} is of order~$\hbar$.
Therefore, the classical Nicolai map is just quadratic in~$g$ and a cubic polynomial in~$A$!
Another way to arrive at this result observes that the cubic and quartic interaction terms in the action~\eqref{lag4}
are separately supersymmetric, allowing us to replace $\tfrac12g^2$ by some new coupling~$\kappa$ in front of the quartic interaction term.
We may now study independent coupling flows in $g$ and~$\kappa$ generated by $R^{(g)}_{g,\kappa}$ and $R^{(\kappa)}_{g,\kappa}$,
respectively, taking into account that the full gaugino propagator has an expansion in powers of both $g$ and~$\kappa$. 
As is outlined in~\cite{LR3}, one can set up Nicolai maps for any path connecting $(0,0)$ to $(g,\kappa)$ in the two-dimensional 
coupling space, but the cancellations observed imply that the classical result is contour independent and simply given by
\begin{equation}
(T_g^{(0)}\!A)(y) \= A(y)\ -\ g\,(R^{(g)}_{0,0}A)(y)\ -\ \kappa\,(R^{(\kappa)}_{0,0}A)(y)\big|_{\hbar=0}\ ,
\end{equation}
in agreement with~\eqref{T4cl}.

Actually, this collapse of the classical map is enforced already by the free-action condition,
because it is linear in the Nicolai map due to $T_g\bA=\bA$. Explicitly,
\begin{equation} \label{fac4}
-\smallint\diff^4x\ T_g^{(0)}\!A\cdot\Box\,\bA 
\= -\smallint\diff^4x\ \bigl\{ A - g\,r_1^{(1)}\!A - \tfrac12 g^2\,r_2^{(2)}\!A\bigr\}\cdot\Box\,\bA 
\= \smallint\diff^4x\ {\cal L}\big|_{\psi=\bps=0} \ ,
\end{equation}
without any contributions of higher order.
It will be interesting to verify the Nicolai-map identities to order~$\hbar$ (determinant matching) and higher~\cite{CLR}.
%

\section{The map in ten dimensions}

\noindent
We now turn to the ten-dimensional super Yang--Mills theory in the light-cone gauge~$A^+{=}\,0$.
Modulo topological terms, its Lagrangian can be cast in the following form~\cite{BGS},
\begin{equation} \label{lag}
\begin{aligned}
{\cal L} \ &=\ \tfrac12 \bigl\{ -A^k\cdot\Box A^k + \im\,\psi\cdot\tfrac{\Box}{\pp}\psi\bigr\} \\[4pt]
&\ +\,g\,\bigl\{ (\pa^k A^k)\cdot\tfrac1{\pp}[(\pp\! A^\ell)\ti A^\ell] - (\pa^k A^\ell)\cdot[A^k\ti A^\ell]
+ \im\,A^k\cdot[\psi\ti\gamma^k\gamma^\ell\tfrac{\pa^\ell}{\pp}\psi] + \im\,(\pa^k A^k)\cdot\tfrac1{\pp}[\psi\ti\psi]\bigr\} \\[4pt]
&\ +\,\tfrac12g^2 \bigl\{ [(\pp\! A^k)\ti A^k +\im\,\psi\ti\psi]\cdot\tfrac1{\pp}\tfrac1{\pp}[(\pp\! A^\ell)\ti A^\ell+\im\,\psi\ti\psi] \\[2pt]
& \qquad\ -\tfrac12[A^k\ti A^\ell]\cdot[A^k\ti A^\ell] -\im\,[A^k\ti\psi\,\gamma^k]\cdot\tfrac1{\pp}[\gamma^\ell\psi\ti A^\ell] \bigr\}\ ,
\end{aligned}
\end{equation}
featuring the transversal components $A^k$ ($k=1,2,\ldots,8$) of the gauge potential and 
an SO(8) Majorana--Weyl spinor gaugino $\psi^{\dot\alpha}$ ($\dot\alpha=1,2,\ldots,8$ suppressed),
both in the adjoint representation of the gauge group, say SU($N$), with gauge coupling~$g$. 
We take the SO(8) spinor metric (or charge conjugation) for this chirality to be $C^{\dot\al\dot\be}=\de^{\dot\al\dot\be}$
and may hence freely raise or lower our spinor indices.
The $8{\times}8$ Weyl matrices $\gamma^k=(\gamma^k_{\alpha\dot\alpha})$ are subject to
\begin{equation}
\gamma^k_{\al\dot\al} \gamma^\ell_{\dot\al\be} + \gamma^\ell_{\al\dot\al} \gamma^k_{\dot\al\be} \= 2\,\de^{k\ell}\de_{\al\be}
\qquad\with \al,\be =1,2,\ldots,8\ ,
\end{equation}
and we shall antisymmetrize their products with weight one, 
{\it e.g.\/} $\ga^{k\ell}=\ga^{[k}\ga^{\ell]}:=\tfrac12(\ga^k\ga^\ell-\ga^\ell\ga^k)$.

The action $S=\smallint\diff^{10}x\,{\cal L}$ is manifestly invariant under half of the original 16 supersymmetries,
\begin{equation} \label{susy}
\de_\al A^i = \im\,(\psi\,\ga^i)_\al \und \de_\al \psi_{\dot\al} = -\ga^i_{\dot\al\al}\,\pp\!A^i\ ,
\end{equation}
which are linearly realized.
Hence, there in a light-cone superspace $(A^i,\psi^{\dot\al})$ can be combined into an SO(8) vector superfield~\cite{BGS},
and there must exist an SO(8) Majorana--Weyl spinor ${\cal M}_\al$ which generates the Lagrangian~\eqref{lag} via
a supervariation,
\begin{equation}
\de_\al\,{\cal M}_\al \= {\cal L} \ + \textrm{total derivative}\ .
\end{equation}
Again, since~\eqref{susy} is linear, $[\tfrac{\pa}{\pa g},\de_\al]=0$ and hence
\begin{equation}
\tfrac{\pa}{\pa g} \smallint\diff^{10}x\ {\cal L}(x) \= \de_\al\smallint\diff^{10}x\ \tfrac{\pa}{\pa g}{\cal M}_\al(x)\ .
\end{equation}

It is not too hard to find this fermionic functional. 
Applying~\eqref{susy} which implies
\begin{equation}
\de_\al(A^i\ti\ga^i\psi)_\al = 8\,((\pp\! A^i)\ti A^i +\im\,\psi\ti\psi) \und
\de_\al((\pp\! A^i)\ti A^i +\im\,\psi\ti\psi) = -\im\,\pp(A^i\ti\psi\,\ga^i)_\al \ ,
\end{equation}
and recalling relations like
\begin{equation}
\psi\ti\ga^i\ga^j\psi' = \psi'\ti\ga^j\ga^i\psi \qquad\textrm{as well as}\qquad
\ga^i\ga^i=8\,\unity \and \ga^i\ga^j\ga^i = -6\ga^j\ ,
\end{equation}
it may be verified directly that~\footnote{
In a general spacetime dimension~$D$, the coefficients work out only when the spinor dimension equals 
the transverse spacetime dimension $D{-}2$, which restricts to the critical values $D=3,4,6$ and~$10$.}
\begin{equation}
\begin{aligned}
{\cal M}_\al \ &=\ \tfrac{1}{16}A^k\cdot\tfrac{\Box}{\pp}(\ga^k\psi)_\al \\[4pt]
&\ +\,g\,\bigl\{ \tfrac{1}{16} A^k\cdot [A^\ell\ti\tfrac{\pa^m}{\pp}(\ga^\ell\ga^k\ga^m\psi)_\al]
- \tfrac18 A^k\cdot \tfrac{\pa^k}{\pp}[A^\ell\ti(\ga^\ell\psi)_\al] \bigr\}\\[4pt]
&\ +\,\tfrac12 g^2 \bigl\{ \tfrac{1}{16} [A^k\ti A^\ell]\cdot\tfrac1{\pp}[A^m\ti(\ga^\ell\ga^k\ga^m\psi)_\al]
+ \tfrac18 [(\pp\! A^k)\ti A^k +\im\,\psi\ti\psi]\cdot\tfrac1{\pp}\tfrac1{\pp}[A^\ell\ti(\ga^\ell\psi)_\al] \bigr\}\ .
\end{aligned}
\end{equation}
With it, the coupling flow operator takes the form
\begin{equation}
\begin{aligned} \label{Rg}
R_g &\= \im\smallint\diff^{10}x\ \< \tfrac{\pa}{\pa g}{\cal M}_\al(x)\ \de_\al \>_\psi
\= -\smallint\diff^{10}x \smallint\diff^{10}y\ \< \tfrac{\pa}{\pa g}{\cal M}_\al(x)\ (\psi\,\ga^i)_\al(y) \>_\psi\ \tfrac{\de}{\de A^i(y)} \\[4pt]
&\= -\smallint\diff^{10}x \smallint\diff^{10}y\ \bigl\{
\tfrac{1}{16} A^k(x)\ti A^\ell(x)\cdot\tfrac{\pa^m}{\pp}\,\tr\,\<\ga^\ell\ga^k\ga^m\psi(x)\,\psi(y)\,\ga^i\>_\psi \cdot \tfrac{\de}{\de A^i(y)}\\
&\qquad\qquad\qquad\qquad\ 
-\tfrac18 A^k(x)\ti\tfrac{\pa^k}{\pp} A^\ell(x) \cdot\,\tr\,\<\ga^\ell\psi(x)\,\psi(y)\,\ga^i\>_\psi\cdot \tfrac{\de}{\de A^i(y)}\\
&\qquad\qquad\qquad\qquad\ 
+\tfrac{g}{16} A^k(x)\ti A^\ell(x)\cdot\tfrac1{\pp}A^m(x)\ti\,\tr\,\<\ga^\ell\ga^k\ga^m\psi(x)\,\psi(y)\,\ga^i\>_\psi \cdot \tfrac{\de}{\de A^i(y)}\\
&\qquad\qquad\qquad\qquad\ 
-\tfrac{g}{8} A^k(x)\ti (\pp\! A^k(x))\cdot\tfrac1{\pp}\tfrac1{\pp}A^\ell(x)\ti\,\tr\,\<\ga^\ell\psi(x)\,\psi(y)\,\ga^i\>_\psi\cdot \tfrac{\de}{\de A^i(y)}\\
&\qquad\qquad\qquad\qquad\ \
+\tfrac{g}{8}\ \< \im\,\psi(x)\ti\psi(x)\cdot\tfrac1{\pp}\tfrac1{\pp}A^\ell(x)\ti\,\tr\,\ga^\ell\psi(x)\,\psi(y)\,\ga^i\>_\psi\cdot \tfrac{\de}{\de A^i(y)}
\bigr\} \\[4pt]
&\= r_1^{(1)}\ +\ g\,(r_2^{(1)}+r_2^{(2)}+r_2^{(3)})\ +\ g^2\,(r_3^{(1)}+r_3^{(2)}+r_3^{(3)})\ +\ \ldots\ ,
\end{aligned}
\end{equation}
where the superscript label~$(1)$ refers to lines 1 and~2, label~$(2)$ to lines 3 and~4, and label~$(3)$ to line~5 of
the expression~\eqref{Rg}.
Again, the color structure is unambiguous without extra brackets. 

Once more, we write down the full gaugino two- and four-point functions to first and zeroth order in~$g$, respectively,
\begin{align} \label{psicorr}
\<\psi^a_{\dot\al}(x)\,\psi^b_{\dot\be}(y)\>_\psi &\= \de^{ab}\de_{\dot\al\dot\be}\tfrac{\pp}{\Box}(x{-}y) \nn \\
&\ +\ g\,f^{acb}\smallint\diff^{10}z\,\tfrac{\pp}{\Box}(x{-}z)\,\bigl\{
\de_{\dot\al\dot\be}\bigl[A^p\tfrac{\pa^p}{\pp}-(\tfrac{\pa^p}{\pp}A^p)\bigr]+
\ga^{pq}_{\dot\al\dot\be}\bigl[A^p\tfrac{\pa^q}{\pp}-\tfrac{\pa^q}{\pp}A^p\bigr] \bigr\}(z)\,
\tfrac{\pp}{\Box}(z{-}y) \nn \\
&\ +\  O(g^2) \\[4pt]
\<(\psi\ti\psi)^c(z)\,\psi^a_{\dot\al}&(x)\,\psi^b_{\dot\be}(y)\>_\psi \= 
-2\,f^{cab}\,\de_{\dot\al\dot\be}\,\tfrac{\pp}{\Box}(z{-}x) \tfrac{\pp}{\Box}(z{-}y) \ + \ O(g)\ , \nn
\end{align}
where we recall that (inverse) derivative actions are limited by round brackets only.
Inserting \eqref{psicorr} into \eqref{Rg} we find
\begin{equation}
r_1^{(1)} \= \tfrac18 \smallint\diff^{10}x \smallint\diff^{10}y\ A^k(x)\ti \bigl\{
\tfrac{\pa^k}{\pp}A^\ell(x)\,\tr(\ga^\ell\ga^i) - \tfrac12 A^\ell(x) \tfrac{\pa^m}{\pp}\,\tr(\ga^\ell\ga^k\ga^m\ga^i)
\bigr\} \cdot \tfrac{\pp}{\Box}(x{-}y)\,\tfrac{\de}{\de A^i(y)} 
\end{equation}
and
\begin{equation}
\begin{aligned}
r_2^{(1)} &\= \tfrac18 \smallint\diff^{10}x \smallint\diff^{10}z \smallint\diff^{10}y\ \bigl\{
A^k(x)\ti \tfrac{\pa^k}{\pp}A^\ell(x)\cdot\tr(\ga^\ell\ga^i)\tfrac{\pp}{\Box}(x{-}z)
\bigl[(A^p\tfrac{\pa^p}{\pp})-\tfrac{\pa^p}{\pp}A^p\bigr](z)  \\
&\qquad\qquad\qquad\qquad\qquad\
+A^k(x)\ti \tfrac{\pa^k}{\pp}A^\ell(x)\cdot\tr(\ga^\ell\ga^{pq}\ga^i)\tfrac{\pp}{\Box}(x{-}z)
\bigl[A^p\tfrac{\pa^q}{\pp}-\tfrac{\pa^q}{\pp}A^p\bigr](z)  \\
&\qquad\qquad\qquad\qquad\qquad\
-\tfrac12 A^k(x)\ti A^\ell(x)\cdot\tr(\ga^\ell\ga^k\ga^m\ga^i)\tfrac{\pa^m}{\Box}(x{-}z)
\bigl[(A^p\tfrac{\pa^p}{\pp})-\tfrac{\pa^p}{\pp}A^p\bigr](z)  \\
&\qquad\qquad\qquad\qquad\qquad\ 
-\tfrac12 A^k(x)\ti A^\ell(x)\cdot\tr(\ga^\ell\ga^k\ga^m\ga^{pq}\ga^i) \tfrac{\pa^m}{\Box}(x{-}z)
\bigl[A^p\tfrac{\pa^q}{\pp}-\tfrac{\pa^q}{\pp}A^p\bigr](z) \bigr\} \\
&\qquad\qquad\qquad\qquad\qquad\qquad
\ti\,\tfrac{\pp}{\Box}(z{-}y)\,\tfrac{\de}{\de A^i(y)} \ , \\[4pt]
r_2^{(2)} &\= \tfrac18 \smallint\diff^{10}x \smallint\diff^{10}y\ \bigl\{
A^k(x)\ti(\pp\! A^k(x))\cdot\tfrac{1}{\pp}\tfrac{1}{\pp}A^\ell(x)\ti\,\tr(\ga^\ell\ga^i)  \\
&\qquad\qquad\qquad\qquad\
-\tfrac12 A^k(x)\ti A^\ell(x)\cdot\tfrac{1}{\pp} A^m(x)\ti\,\tr(\ga^\ell\ga^k\ga^m\ga^i) \bigr\}
\tfrac{\pp}{\Box}(x{-}y)\,\tfrac{\de}{\de A^i(y)} \ ,\\[4pt]
r_2^{(3)} &\= \tfrac{\im}{4} \smallint\diff^{10}x \smallint\diff^{10}y\ 
f^{cab}f^{cda}\,\tfrac{1}{\pp}\tfrac{1}{\pp} A_d^\ell(x)\,\tr(\ga^\ell\ga^i)
\tfrac{\pp}{\Box}(z{-}x)\tfrac{\pp}{\Box}(z{-}y)\big|_{z=x} \,\tfrac{\de}{\de A_b^i(y)} \ .
\end{aligned}
\end{equation}
Evaluating the spin traces with the help of
\begin{equation}
\begin{aligned}
&\ga^{\ell k}\ga^m\=\de^{mk}\ga^\ell-\de^{m\ell}\ga^k+\ga^{\ell km} \qquad\textrm{and}\qquad
\tr(\ga^\ell\ga^i)=8\,\de^{i\ell}\ ,\quad
\tr(\ga^{\ell km}\ga^i)=0\ ,\\
&\tr(\ga^{\ell}\ga^{pq}\ga^i)=8\,(\de^{iq}\de^{\ell p}{-}\de^{ip}\de^{\ell q})\ ,\quad
\tr(\ga^{\ell km}\ga^{pq}\ga^i)=16\,(\de^{mi}\de^{p[k}\de^{\ell]q}+\de^{mp}\de^{q[k}\de^{\ell]i}+\de^{mq}\de^{i[k}\de^{\ell]p})
\end{aligned}
\end{equation}
we arrive at
\begin{equation}
\begin{aligned}
r_1^{(1)} &\= \smallint\diff^{10}x  \smallint\diff^{10}y\ A^k(x)\ti \bigl\{
\tfrac{\pa^k}{\pp}A^i(x)-A^i(x)\tfrac{\pa^k}{\pp}\bigr\} \tfrac{\pp}{\Box}(x{-}y)\cdot\,\tfrac{\de}{\de A^i(y)} \\[2pt]
&\= \smallint\diff^{10}x  \smallint\diff^{10}y\ A^i(x)\ti \bigl\{
A^k(x)\tfrac{\pa^k}{\pp}-(\tfrac{\pa^k}{\pp}A^k(x))\bigr\}\tfrac{\pp}{\Box}(x{-}y)\cdot\,\tfrac{\de}{\de A^i(y)}
\end{aligned}
\end{equation}
and (splitting $(1)$ into $(1a)$ and $(1b)$ and making $\hbar$ explicit)
\begin{align}
r_2^{(1a)} &\= \smallint\diff^{10}x \smallint\diff^{10}z \smallint\diff^{10}y\ \bigl\{
A^k\ti\bigl[\tfrac{\pa^k}{\pp}A^i-A^i\tfrac{\pa^k}{\pp}\bigr](x)\cdot \tfrac{\pp}{\Box}(x{-}z)
\bigl[ A^\ell\tfrac{\pa^\ell}{\pp}-(\tfrac{\pa^\ell}{\pp}A^\ell)\bigr](z) \nn \\
&\qquad\qquad\qquad\qquad\qquad
+A^k\ti\bigl[\tfrac{\pa^k}{\pp}A^\ell-A^\ell\tfrac{\pa^k}{\pp}\bigr](x)\cdot\tfrac{\pp}{\Box}(x{-}z)
\bigl[ A^\ell\tfrac{\pa^i}{\pp}-\tfrac{\pa^i}{\pp}A^\ell-A^i\tfrac{\pa^\ell}{\pp}+\tfrac{\pa^\ell}{\pp}A^i\bigr](z) \bigr\} \nn \\
&\qquad\qquad\qquad\qquad\qquad\qquad
\ti\,\tfrac{\pp}{\Box}(z{-}y)\,\tfrac{\de}{\de A^i(y)} \ , \nn \\[4pt]
r_2^{(1b)} &\= \smallint\diff^{10}x \smallint\diff^{10}z \smallint\diff^{10}y\ 
\bigl[ A^k\ti A^\ell\,\pa^i+A^\ell\ti A^i\,\pa^k+A^i\ti A^k\,\pa^\ell\bigr](x)\cdot\tfrac{1}{\Box}(x{-}z)
\bigl[A^\ell\tfrac{\pa^k}{\pp}-\tfrac{\pa^k}{\pp}A^\ell\bigr](z) \nn \\
&\qquad\qquad\qquad\qquad\qquad\qquad
\ti\,\tfrac{\pp}{\Box}(z{-}y)\,\tfrac{\de}{\de A^i(y)} \ ,\\[4pt]
r_2^{(2)} &\= \smallint\diff^{10}x \smallint\diff^{10}y\ \bigl\{
A^k\ti(\pp\! A^k)\cdot\tfrac{1}{\pp}\tfrac{1}{\pp}A^i-A^k\ti A^i\cdot\tfrac{1}{\pp} A^k
\bigr\}(x) \tfrac{\pp}{\Box}(x{-}y)\ti\,\tfrac{\de}{\de A^i(y)} \ , \nn \\[4pt]
r_2^{(3)} &\= -2N\,\im\hbar\,\tfrac{1}{\pp}\tfrac{1}{\pp} A^i(x)\cdot\tfrac{\pp}{\Box}(z{-}x)\tfrac{\pp}{\Box}(z{-}y)\big|_{z=x}
\,\tfrac{\de}{\de A^i(y)} \ . \nn
\end{align}

Again, we are in a position to write down the Nicolai map~\eqref{Tg4} to order~$g^2$
and observe that $r_1^{(1)}r_1^{(1)}\!A$ cancels two thirds of the terms in~$r_2^{(1a)}\!A$. 
In contrast to the situation in four dimensions, one third of $r_2^{(1a)}\!A$ and all of $r_2^{(1b)}\!A$ remain.
The final expression simplifies to
\begin{align} \label{T10}
(T_gA)^i(y) &\= A^i(y)\ -\ g\, \smallint\diff^{10}x\ A^k(x)\ti \bigl\{
\tfrac{\pa^k}{\pp}A^i-A^i\tfrac{\pa^k}{\pp}\bigr\}(x) \tfrac{\pp}{\Box}(x{-}y) \nn\\[2pt]
-\ \tfrac12 g^2 \smallint&\diff^{10}x \smallint\diff^{10}z\
A^k\ti\bigl\{ \tfrac{\pa^k}{\pp}A^\ell-A^\ell\tfrac{\pa^k}{\pp} \bigr\}(x) \ti \tfrac{\pp}{\Box}(x{-}z)
\bigl\{ A^\ell\tfrac{\pa^i}{\pp}-\tfrac{\pa^i}{\pp}A^\ell\bigr\}(z) \tfrac{\pp}{\Box}(z{-}y) \nn\\
-\ \tfrac12 g^2 \smallint&\diff^{10}x \smallint\diff^{10}z\ \bigl\{
A^k\ti A^\ell\ti\pa^i +  A^\ell\ti A^i\ti\pa^k  + A^i\ti A^k\ti\pa^\ell \bigr\}(x)
\tfrac{1}{\Box}(x{-}z) \bigl\{A^\ell\tfrac{\pa^k}{\pp}-\tfrac{\pa^k}{\pp}A^\ell\bigr\}(z) \tfrac{\pp}{\Box}(z{-}y) \nn\\
&\ \ -\ \tfrac12 g^2 \smallint\diff^{10}x\ A^k\ti \bigl\{
(\pp\!A^k)\ti\tfrac{1}{\pp}\tfrac{1}{\pp}A^i-A^i\ti\tfrac{1}{\pp} A^k\bigr\}(x)\tfrac{\pp}{\Box}(x{-}y) \\
&\ \ +\ \im\hbar N g^2 \smallint\diff^{10}x\ \tfrac{1}{\pp}\tfrac{1}{\pp} A^i(x)
\tfrac{\pp}{\Box}(z{-}x)\tfrac{\pp}{\Box}(z{-}y)\big|_{z=x} \ +\ \ O(g^3) \nn\\[2pt]
&\ =:\  A^i(y)\ +\ g\,(t_1 A)^i(y)\ +\ g^2 \bigl(
t_2^{(1a)}\!A + t_2^{(1b)}\!A + t_2^{(2)}\!A + t_2^{(3)}\!A \bigr){}^i(y) \ +\  O(g^3) \ .\nn
\end{align}
The first, fourth and fifth line follow directly from the structure of~$\cal M$ by taking the gaugino propagator to be the free one.
The second and third line (those with double integrals) are remainders of the expansion of the full gaugino propagators.
Interestingly, the third line, coming from $r_2^{(1b)}\!A$ and being fully antisymmetric in $[k\ell i]$, 
repeats the structure of the covariant map in the Landau gauge~\cite{Nic2}, but will be absent in four spacetime 
({\it i.e.} two transverse) dimensions.
Lines~1--4 are part of the classical $T_g^{(0)}\!A$, while line~5 is the leading part $g^2\,t_2^{(3)}\!A$ 
of the quantum correction~$\hbar\,T_g^{(1)}\!A$.

Let us check the free-action condition
\begin{equation} \label{fac}
-\tfrac12\smallint\diff^{10}x\ (T_g^{(0)}\!A)^i\cdot\Box\,(T_g^{(0)}\!A)^i 
\ \buildrel{!}\over{=} \ \smallint\diff^{10}x\ {\cal L}\big|_{\psi=0}
\end{equation}
to second order in the coupling. Surprisingly, the right-hand side is again saturated already by the classical contribution 
linear in the map and ignoring the $g$~expansion of the gaugino propagator,
\begin{equation}
-\tfrac12\smallint\diff^{10}x\ \bigl\{ A\ +\ 2g\,t_1 A\ +\ 2g^2 t_2^{(2)}\!A \bigr\}{}^i \cdot\Box\,A^i 
\= \smallint\diff^{10}x\ {\cal L}\big|_{\psi=0}\ .
\end{equation}
It follows that, besides the vanishing of all higher orders on the left-hand side of~\eqref{fac},
the remaining classical contributions at order~$g^2$ must cancel out,
\begin{equation}
\begin{aligned}
0 &\ \buildrel{!}\over{=} \ \smallint\diff^{10}x\ \bigl\{
(t_1 A)^i\cdot\Box\,(t_1 A)^i + 2\,(t_2^{(1a)}\!A+t_2^{(1b)}\!A)^i \cdot \Box\,A^i \bigr\} \\[4pt]
&\= \smallint\diff^{10}x \smallint\diff^{10}y\ \Bigl\{
A^k\ti\bigl[\tfrac{\pa^k}{\pp}A^i-A^i\tfrac{\pa^k}{\pp}\bigr](x)\cdot\tfrac{\pp\overleftarrow{\pp}}{\Box}(x{-}y)\,
\bigl[\overleftarrow{\tfrac{\pa^\ell}{\pp}}A^i - A^i\overleftarrow{\tfrac{\pa^\ell}{\pp}}\bigr]\ti A^\ell(y) \\
&\qquad\qquad\qquad\quad
-\,A^k\ti\bigl[\tfrac{\pa^k}{\pp}A^i-A^i\tfrac{\pa^k}{\pp}\bigr](x)\cdot\tfrac{\pp}{\Box}(x{-}y)\,
\bigl[A^i\tfrac{\pa^\ell}{\pp}-\tfrac{\pa^\ell}{\pp}A^i\bigr]\ti\,\pp\!A^\ell(y) \\[2pt]
&\qquad\qquad\qquad\quad
+\,\bigl\{A^k\ti A^\ell\,\pa^i +  A^\ell\ti A^i\,\pa^k  + A^i\ti A^k\,\pa^\ell\bigr\}(x)
\cdot\tfrac{1}{\Box}(x{-}y)\,\bigl[A^k\tfrac{\pa^\ell}{\pp}-\tfrac{\pa^\ell}{\pp}A^k\bigr]\ti\,\pp\!A^i(y) \\[4pt]
&\= \tfrac12 \smallint\diff^{10}x \smallint\diff^{10}y\ \Bigl\{
A^k\ti A^\ell\,\pa^i(x)\tfrac{1}{\Box}(x{-}y)\bigl[ \pa^\ell(A^k\ti A^i) - \tfrac{\pa^\ell}{\pp}\pp\!(A^k\ti A^\ell)\bigr](y) \\
&\qquad\qquad\qquad\quad
+\,\tfrac12\,A^i\ti A^k\,\pa^\ell(x)\tfrac{1}{\Box}(x{-}y)\bigl[\pa^\ell(A^k\ti A^i)-\tfrac{\pa^\ell}{\pp}\pp\!(A^k\ti A^i)\bigr](y)
\Bigr\} \\[2pt]
&\= 0 \ .\qquad \checkmark 
\end{aligned}
\end{equation}
Interestingly, the $t_2^{(1a)}$ part cancels against the $t_1 t_1$ contribution, while the `covariant' piece $t_2^{(1b)}$
vanishes by itself due to the difference inside the last square bracket.
This verifies the Nicolai map~\eqref{T10} to second order in the gauge coupling, 
and it supports the accuracy of the construction, which thus may be extended to any higher order.
As for the four-dimensional map, it may be instructive to check the higher-order (quantum) Nicolai-map identities~\cite{CLR},
although the construction guarantees their validity.

\subsection*{Acknowledgments}
\noindent
We are grateful to Hermann Nicolai for pointing out the light-cone formulation and for fruitful discussions
at an early stage of this work. We also thank Lorenzo Casarin for useful conversations.

\newpage



\begin{thebibliography}{99}
\addtolength{\itemsep}{-1pt}

\bibitem{Nic1}
H.~Nicolai,
{\it On a new characterization of scalar supersymmetric theories},
\href{https://dx.doi.org/10.1016/0370-2693(80)90138-0}
{{\it Phys.\ Lett.\ B} {\bf 89} (1980) 341}.
%
\bibitem{Nic2}
H.~Nicolai,
{\it Supersymmetry and functional integration measures},
\href{https://dx.doi.org/10.1016/0550-3213(80)90460-5}
{{\it Nucl.\ Phys.\ B} {\bf176} (1980) 419}.
%
\bibitem{Nic3}
H.~Nicolai,
{\it Supersymmetric functional integration measures},\\
lectures delivered at the NATO Advanced Study Institute on Supersymmetry,\\
Bonn, Germany, 20--31 Aug 1984,
\href{https://cds.cern.ch/record/155731?ln=en}
{pp.393--420, eds. K.~Dietz et. al., {\it Plenum Press} (1984)}.
%
\bibitem{Lprague}
O.~Lechtenfeld,
{\it The Nicolai-map approach to supersymmetry},\\
Talk at QTS12, Prague, 24-28 July 2023,
{}[\href{https://arxiv.org/abs/2309.00481}{arXiv:2309.00481 [hep-th]}].
%
\bibitem{CLR}
L.~Casarin, O.~Lechtenfeld and M.~Rupprecht,
{\it Nicolai maps with four-fermion interactions},\\
\href{https://doi.org/10.1007/JHEP12(2023)132}
{{\it JHEP} {\bf 12} (2023) 132}
{}[\href{https://doi.org/10.48550/arXiv.2310.19946}{arXiv:2310.19946 [hep-th]}].
%
%
\bibitem{DL1}
K.~Dietz and O.~Lechtenfeld,
{\it Nicolai maps and stochastic observables from a coupling constant flow},\\
\href{https://dx.doi.org/10.1016/0550-3213(85)90132-4}
{{\it Nucl.\ Phys.\ B} {\bf 255} (1985) 149}.
%
\bibitem{L1}
O.~Lechtenfeld,
{\it Construction of the Nicolai mapping in supersymmetric field theories},\\
Ph.D.\ Thesis, Bonn University (1984),
\href{https://lib-extopc.kek.jp/preprints/PDF/2000/0030/0030157.pdf}
{internal report {\it BONN-IR-84-42}, ISSN-0172-8741}.
%
\bibitem{L2}
O.~Lechtenfeld,
{\it Stochastic variables in ten dimensions?},
\href{https://doi.org/10.1016/0550-3213(86)90531-6}
{{\it Nucl.\ Phys.\ B} {\bf 274} (1986) 633}.
%
\bibitem{LR1}
O.~Lechtenfeld and M.~Rupprecht,
{\it Universal form of the Nicolai map},\\
\href{https://doi.org/10.1103/PhysRevD.104.L021701}
{{\it Phys.\ Rev.\ D} {\bf 104} (2021) L021701}
[\href{https://arxiv.org/abs/2104.00012}{arXiv:2104.00012 [hep-th]}].
%
\bibitem{ALMNPP}
S.~Ananth, O.~Lechtenfeld, H.~Malcha, H.~Nicolai, C.~Pandey and S.~Pant,
{\it Perturbative linearization of supersymmetric Yang--Mills theory},
\href{https://dx.doi.org/10.1007/JHEP10(2020)199}
{{\it JHEP} {\bf 10} (2020) 199}
[\href{https://arxiv.org/abs/2005.12324}{arXiv:2005.12324 [hep-th]}].
%
\bibitem{MN}
H.~Malcha and H.~Nicolai,
{\it Perturbative linearization of super-Yang--Mills
        theories in general gauges},\\
\href{https://doi.org/10.1007/JHEP06(2021)001}
{{\it JHEP} {\bf 06} (2021) 001}
[\href{https://arxiv.org/abs/2104.06017}{arXiv:2104.06017 [hep-th]}].
%
\bibitem{LR2}
O.~Lechtenfeld and M.~Rupprecht,
{\it Construction method for the Nicolai map in supersymmetric Yang--Mills theories},
\href{https://doi.org/10.1016/j.physletb.2021.136413}
{{\it Phys.\ Lett.\ B} {\bf 819} (2021) 136413}
[\href{https://arxiv.org/abs/2104.09654}{arXiv:2104.09654 [hep-th]}].
%
\bibitem{LR4}
O.~Lechtenfeld and M.~Rupprecht,
{\it An improved Nicolai map for super Yang--Mills theory},\\
\href{https://doi.org/10.1016/j.physletb.2023.137681}
{{\it Phys.\ Lett.\ B} {\bf 838} (2023) 137681}
[\href{https://arxiv.org/abs/2211.07660}{arXiv:2211.07660 [hep-th]}].
%
\bibitem{HRT}
Y.~Hassoun, A.~Restuccia and J.G.~Taylor,
{\it The reduction of N=1 supersymmetric Yang--Mills theory to the light-cone gauge},
\href{https://doi.org/10.1016/0370-2693(83)91434-X}
{{\it Phys.\ Lett.\ B} {\bf 124} (1983) 197}.
%
\bibitem{BGS}
L.~Brink, M.B.~Green and J.H.~Schwarz,
{\it Ten-dimensional supersymmetric Yang--Mills theory with SO(8)-covariant light-cone superfields},
\href{https://doi.org/10.1016/0550-3213(83)90096-2}
{{\it Nucl.\ Phys.\ B} {\bf 223} (1983) 125}.
%
\bibitem{dAFFV}
V.~de Alfaro, S.~Fubini, G.~Furlan and G.~Veneziano,
{\it Stochastic identities in quantum theory},\\
\href{https://doi.org/10.1016/0550-3213(85)90127-0}
{{\it Nucl.\ Phys.\ B} {\bf 255} (1985) 1}.
%
\bibitem{LR3}
O.~Lechtenfeld and M.~Rupprecht,
{\it Is the Nicolai map unique?},\\
\href{https://doi.org/10.1007/JHEP09(2022)139}
{{\it JHEP} {\bf 09} (2022) 139}
[\href{https://arxiv.org/abs/2207.09471}{arXiv:2207.09471 [hep-th]}].
%
\bibitem{BLN}
L.~Brink, O.~Lindgren and E.W.~Nilsson,
{\it N=4 Yang--Mills theory on the light cone},\\
\href{https://doi.org/10.1016/0550-3213(83)90678-8}
{{\it Nucl.\ Phys.\ B} {\bf 212} (1983) 401}.
%
\bibitem{Man}
S.~Mandelstam,
{\it Light-cone superspace and ultraviolet finiteness of the N=4 model},\\
\href{https://doi.org/10.1016/0550-3213(83)90179-7}
{{\it Nucl.\ Phys.\ B} {\bf 213} (1983) 149}.
%
\bibitem{DL2}
K.~Dietz and O.~Lechtenfeld,
{\it Ghost-free quantisation of non-Abelian gauge theories via the Nicolai transformation 
	of their supersymmetric extensions},
\href{https://doi.org/10.1016/0550-3213(85)90642-X}
{{\it Nucl.\ Phys.\ B} {\bf 259} (1985) 397}.
%
\bibitem{NP}
H.~Nicolai and J.~Plefka,
{\it ${\cal N}{=}\,4$ super-Yang--Mills correlators without anticommuting variables},\\
\href{https://dx.doi.org/10.1103/PhysRevD.101.125013}
{{\it Phys.\ Rev.\ D} {\bf 101} (2020) 125013}
[\href{https://arxiv.org/abs/2003.14325}{arXiv:2003.14325 [hep-th]}].
%
\bibitem{Mal}
H.~Malcha,
{\it Two loop ghost free quantisation of Wilson loops in ${\cal N}{=}\,4$ supersymmetric Yang--Mills},\\
\href{https://doi.org/10.1016/j.physletb.2022.137377}
{{\it Phys.\ Lett.\ B} {\bf 833} (2022) 137377}
[\href{https://arxiv.org/abs/2206.02919}{arXiv:2206.02919 [hep-th]}].
%
\end{thebibliography}
\end{document}